\documentclass[pdflatex,sn-mathphys-num]{sn-jnl}

\usepackage{hyperref}
\usepackage{tabularx}
\usepackage{array}
\usepackage{subcaption}

\usepackage{graphicx}%
\usepackage{multirow}%
\usepackage{amsmath,amssymb,amsfonts}%
\usepackage{amsthm}%
\usepackage{mathrsfs}%
\usepackage[title]{appendix}%
\usepackage{xcolor}%
\usepackage{textcomp}%
\usepackage{manyfoot}%
\usepackage{booktabs}%
\usepackage{algorithm}%
\usepackage{algorithmicx}%
\usepackage{algpseudocode}%
\usepackage{listings}%

\usepackage{makecell}

\newcolumntype{\perftable}{p{0.13\linewidth} p{0.15\linewidth} p{0.15\linewidth} p{0.15\linewidth} p{0.15\linewidth} p{0.15\linewidth}}
\newcolumntype{\localizationtable}{p{0.2\linewidth} p{0.15\linewidth} p{0.15\linewidth} p{0.15\linewidth}}
\newcolumntype{\datasettable}{p{0.30\linewidth} p{0.16\linewidth} p{0.16\linewidth} p{0.16\linewidth}}

\newcommand{\cancerexplain}{The Cancer category represents overall detection performance, while Calcification, Mass, and Other categories show detailed performance metrics for each specific lesion type.}

\begin{document}

\title[Deep Learning-Based Breast Cancer Detection in Mammography: A Multi-Center Validation Study in Thai Population]{Deep Learning-Based Breast Cancer Detection in Mammography: A Multi-Center Validation Study in Thai Population}



\author*[1]{\fnm{Isarun} \sur{Chamveha}}\email{isarun@perceptra.tech}
\equalcont{These authors contributed equally to this work.}
\author[1]{\fnm{Supphanut} \sur{Chaiyungyuen}}
\equalcont{These authors contributed equally to this work.}
\author[1]{\fnm{Sasinun} \sur{Worakriangkrai}}
\author[1]{\fnm{Nattawadee} \sur{Prasawang}}
\author[2]{\fnm{Warasinee} \sur{Chaisangmongkon}}

\author[3,4]{\fnm{Pornpim} \sur{Korpraphong}}
\author[3,4]{\fnm{Voraparee} \sur{Suvannarerg}}
\author[3,4]{\fnm{Shanigarn} \sur{Thiravit}}
\author[4]{\fnm{Chalermdej} \sur{Kannawat}}
\author[5]{\fnm{Kewalin} \sur{Rungsinaporn}}
\author[5]{\fnm{Suwara} \sur{Issaragrisil}}
\author[6]{\fnm{Payia} \sur{Chadbunchachai}}
\author[6]{\fnm{Pattiya} \sur{Gatechumpol}}
\author[6]{\fnm{Chawiporn} \sur{Muktabhant}}
\author[6]{\fnm{Patarachai} \sur{Sereerat}}

\affil*[1]{\orgname{Perceptra Co., Ltd}, \orgaddress{\city{Bangkok}, \country{Thailand}}}
\affil*[2]{\orgdiv{Institute of Field Robotics}, \orgname{King Mongkut's University of Technology Thonburi}, \orgaddress{\city{Bangkok}, \country{Thailand}}}
\affil*[3]{\orgdiv{Department of Radiology, Faculty of Medicine Siriraj Hospital}, \orgname{Mahidol University}, \orgaddress{\city{Bangkok}, \country{Thailand}}}
\affil*[4]{\orgdiv{Thanyarak Breast Center}, \orgname{Siriraj Hospital}, \orgaddress{\city{Bangkok}, \country{Thailand}}}
\affil*[5]{\orgname{Bangkok Hospital}, \orgaddress{\city{Bangkok}, \country{Thailand}}}
\affil*[6]{\orgdiv{Department of Radiology, Faculty of Medicine}, \orgname{Khonkaen University}, \orgaddress{\city{Khonkaen}, \country{Thailand}}}

%
%
%
%
%
%

\abstract{
This study presents a deep learning system for breast cancer detection in mammography, developed using a modified EfficientNetV2 architecture with enhanced attention mechanisms. The model was trained on mammograms from a major Thai medical center and validated on three distinct datasets: an in-domain test set (9,421 cases), a biopsy-confirmed set (883 cases), and an out-of-domain generalizability set (761 cases) collected from two different hospitals. For cancer detection, the model achieved AUROCs of 0.89, 0.96, and 0.94 on the respective datasets. The system's lesion localization capability, evaluated using metrics including Lesion Localization Fraction (LLF) and Non-Lesion Localization Fraction (NLF), demonstrated robust performance in identifying suspicious regions. Clinical validation through concordance tests showed strong agreement with radiologists: 83.5\% classification and 84.0\% localization concordance for biopsy-confirmed cases, and 78.1\% classification and 79.6\% localization concordance for out-of-domain cases. Expert radiologists' acceptance rate also averaged 96.7\% for biopsy-confirmed cases, and 89.3\% for out-of-domain cases. The system achieved a System Usability Scale score of 74.17 for source hospital, and 69.20 for validation hospitals, indicating good clinical acceptance. These results demonstrate the model's effectiveness in assisting mammogram interpretation, with the potential to enhance breast cancer screening workflows in clinical practice.}

\keywords{Breast Cancer, Mammography, Screening, Deep Learning, Artificial Intelligence, Multi-Center Study}

\maketitle

\section{Introduction}
In 2020, breast cancer was the most frequently diagnosed cancer type globally, with over 2.26 million new cases reported \cite{who-report}. By 2022, the number of new breast cancer cases had risen to more than 2.31 million, making it the second most common cancer type after lung cancer. Mammography screening serves as the primary tool for early breast cancer detection worldwide, demonstrating its effectiveness in reducing breast cancer mortality through early detection. However, the increasing volume of screening mammograms coupled with a shortage of expert radiologists presents a significant challenge to healthcare systems, particularly in developing countries.

Recent advances in artificial intelligence (AI), particularly deep learning, have shown promising results in medical image analysis. These algorithms can process large volumes of data and learn complex patterns in images, enabling accurate detection of cancerous lesions even at early stages \cite{martinez2023deep}. Several studies have demonstrated the potential of deep learning in breast cancer detection. Research from South Korea \cite{kim2018applying} and the United States \cite{teare2017malignancy} has shown that deep learning-based systems can accurately detect cancerous lesions in mammograms, leading to regulatory approvals from organizations such as the FDA \cite{zebra-medical} and CE Mark \cite{lunit-insight}.

However, the clinical implementation of AI systems for mammogram interpretation faces several challenges. A primary concern is the limited generalizability of deep learning models across different populations and imaging protocols. Models trained on datasets from one region or population may not perform adequately on others due to variations in imaging protocols, equipment, and demographic characteristics \cite{guan2020analysis}. This challenge is particularly relevant in Southeast Asian populations, where breast tissue density patterns and cancer presentation may differ from Western populations where most AI systems are developed and validated.

Additionally, rigorous clinical validation is essential to ensure the reliability and effectiveness of AI systems in real-world clinical settings. While several studies have reported promising results in controlled environments, comprehensive validation studies examining model performance across different clinical scenarios, patient populations, and healthcare settings are limited. This gap is particularly notable in the context of Thai population, where no extensive validation study has been conducted to date.

This study aims to address these challenges by:
\begin{enumerate}
    \item Developing and validating a deep learning system for breast cancer detection in mammography, specifically trained and tested on Thai population data
    \item Evaluating the model's performance on multiple independent datasets, including biopsy-confirmed cases and out-of-domain data from different hospitals
    \item Assessing the clinical utility and user acceptance of the system through comprehensive usability studies with practicing radiologists
\end{enumerate}

We present a modified EfficientNetV2-based architecture trained on a large dataset from a major Thai medical center. The model is designed to detect and localize four major types of breast lesions: masses, calcifications, axillary adenopathy, and architectural distortions. Our validation approach encompasses both technical performance metrics and clinical usability assessments, providing a comprehensive evaluation of the system's potential for clinical implementation.

This work contributes to the field by:
\begin{itemize}
    \item Providing the first large-scale validation study of an AI system for mammography interpretation in Thai population
    \item Demonstrating a comprehensive evaluation framework that combines technical performance metrics with clinical usability assessments
    \item Addressing the generalizability challenge through validation on multiple independent datasets
    \item Offering insights into the practical implementation of AI systems in clinical mammography workflows
\end{itemize}

\section{Deep Learning for Mammogram Analysis}
Breast cancer remains a significant global health concern, with over 2.31 million new cases reported in 2022, making it the second most common cancer type after lung cancer. The increasing prevalence of breast cancer, coupled with limited availability of expert medical personnel, has created a pressing need for optimizing cancer screening workflows. Mammography, as the most accessible and cost-effective screening method, plays a crucial role in early detection, but interpretation of mammograms presents several challenges that artificial intelligence (AI) could potentially address.

Over the past decade, researchers have investigated computer-aided mammography analysis extensively \cite{yoon2021deep}. A study \cite{kim2020changes} reported an area under the receiver operating characteristic curve (AUROC) of 0.959 for their deep learning system in detecting malignant lesions, with a sensitivity of 91.5\% and specificity of 88.5\%, comparable to expert radiologists who achieved an average AUROC of 0.965. Their approach involved training the ResNet-34 architecture on both pixel-level datasets indicating lesion locations and image-level mammogram datasets, demonstrating impressive performance in detecting cancerous masses, distortions, and asymmetries in dense breasts. In a large-scale study \cite{schaffter2020evaluation}, researchers used 144,231 screening mammograms (including 952 cancer-positive cases) and validated on an independent cohort of 166,578 examinations. Notably, integrating AI with radiologist assessment improved overall accuracy, achieving an AUROC of 0.942 and reducing false-positive rates by 1.2\%.

Technical approaches have evolved from simple convolutional neural networks to more sophisticated architectures. The EfficientNet family of models has shown particular promise, with EfficientNetV2 achieving significant improvements in both accuracy and computational efficiency \cite{tan2021efficientnetv2}. Another study \cite{kim2018applying} demonstrated that using data-driven imaging biomarkers in mammography screening could achieve high accuracy while maintaining clinical relevance. Various approaches have been employed to detect cancer in mammographic images, with some studies analyzing each mammographic view independently while others utilized all four standard views (left CC, left MLO, right CC, and right MLO). Including breast density information has been shown to improve the accuracy of predicting breast cancer risk \cite{brentnall2015mammographic}.

Despite promising results in clinical validation studies, performance metrics often vary significantly across different populations and imaging protocols, highlighting the challenge of generalizability \cite{guan2020analysis}. The evaluation landscape has expanded beyond simple classification metrics to include localization accuracy and region-based metrics, with state-of-the-art systems achieving high Lesion Localization Fraction values while maintaining low Non-lesion Localization Fraction rates. Recent work has focused on developing models that detect and localize four major lesions: mass, calcification, axillary adenopathy, and architectural distortion using colormap visualizations. These comprehensive approaches demonstrate both the progress made in AI-based mammography analysis and the remaining challenges in creating systems that can be effectively integrated into clinical workflows while maintaining consistent performance across diverse populations.

\section{Material \& Method}
In this section, we explain the dataset and methodology used for development and evaluation of our breast cancer detection algorithm.

\subsection{Development Dataset}
We collected 88,347 full-field digital mammography (FFDM) images and corresponding radiologists reports taken from 2007 to 2021 at the Thanyarak Breast Center, Siriraj Hospital, Bangkok. Each mammography comprises two standard views of each breast: bilateral craniocaudal (CC) and mediolateral oblique (MLO). Each report contains radiological findings from both mammograms and ultrasound along with their final BIRADS category.


This model targets abnormalities in 2D mammography images, so we include only cases with breast lesions visible in mammograms and exclude those with lesions found in ultrasound reports without corresponding mammographic findings. We also exclude images containing breast implants, pacemaker devices, poor image quality, or non-standard views.

\begin{figure}[htbp]
    \centering
    \includegraphics[width=\linewidth]{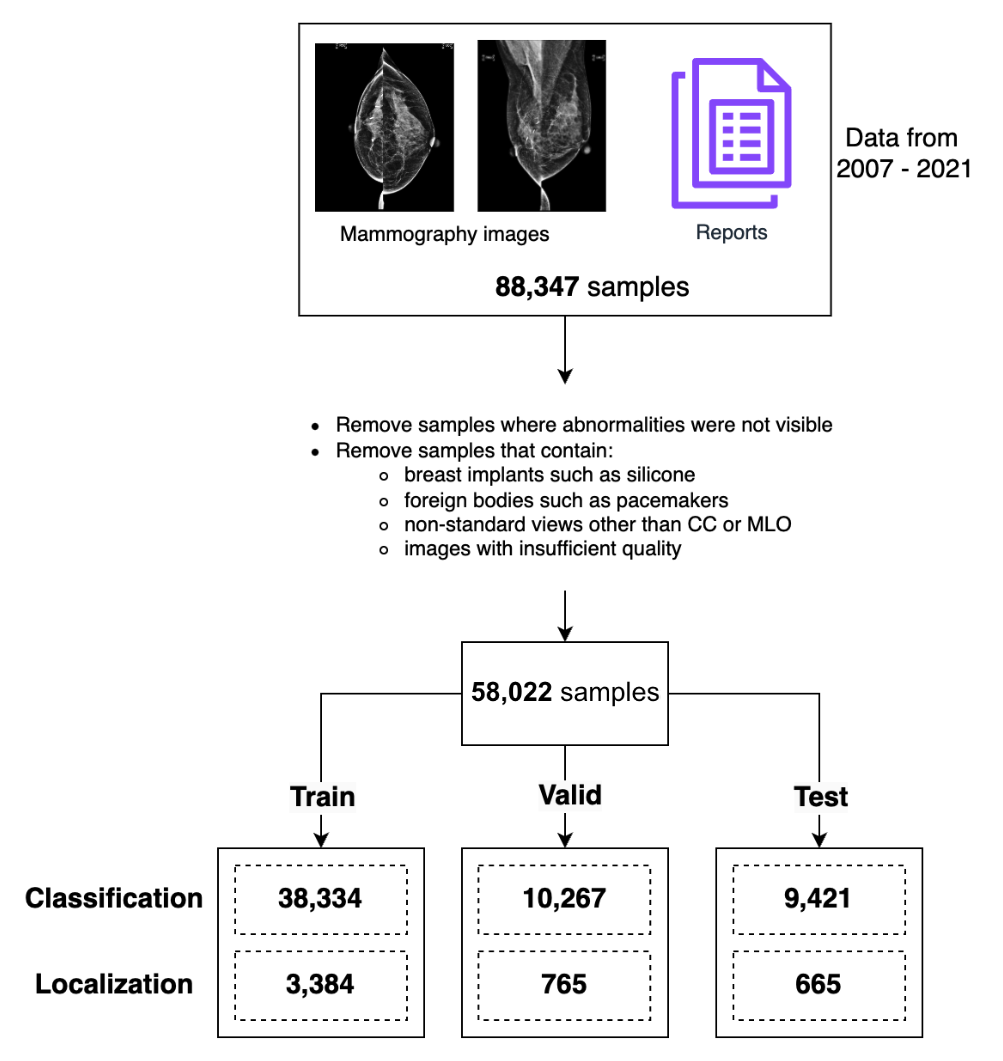}
    \caption{MMG development dataset.}
    \label{fig:mmg-development}
\end{figure}

Our final dataset contains 58,022 mammograms which are divided into train, valid and test set in approximately 0.70-0.15-0.15 ratio. We also curated a localization dataset for a subset of the development dataset. The localization set contains 4,814 (3,384 train, 765 valid, 665 test) samples as shown in Figure \ref{fig:mmg-development}. The distribution of the dataset based on various categories is shown in Table \ref{tab:dataset-distribution}.

\begin{table}[htbp]
    \centering
    \caption{Data Distribution of the development dataset.}
    \label{tab:dataset-distribution}
    \begin{tabular}{lc}
        \midrule
        \textbf{Condition} & \textbf{\# cases} \\
        \midrule
        Benign Calcification & 27,203 \\
        Calcification & 5,914 \\
        Benign Mass & 4,129 \\
        Mass & 3,747 \\
        Normal & 24,925 \\
        Benign Axillary adenopathy & 543 \\
        Axillary adenopathy & 2,054 \\
        Benign Architectural distortion & 327 \\
        Architectural distortion & 1,662 \\
        \midrule
        \textbf{BIRADS Category} & \textbf{\# cases} \\
        \midrule
        1 & 24,925 \\
        2 & 21,975 \\
        3 & 3,718 \\
        4 & 504 \\
        4A & 1,357 \\
        4B & 2,804 \\
        4C & 1,210 \\
        5 & 1,135 \\
        6 & 394 \\
        \midrule
        \textbf{Breast Density} & \textbf{\# cases} \\
        \midrule
        Almost entirely fat & 1,253 \\
        Scattered areas of fibroglandular densities & 14,768 \\
        Heterogeneously dense & 39,008 \\
        Extremely dense & 2,993 \\
        \bottomrule
    \end{tabular}
\end{table}

\subsection{Establishing reference standard}
To establish reference standards, mammogram labels were extracted from a database containing diagnostic assessments made by radiologists who reviewed the mammograms. For a subset of the dataset, we collected localization ground truths for suspicious and benign lesions based on report specifications. These localization labels were further verified by expert radiologists prior to model development.

Since labeling lesions at pixel-level is both highly expensive and time-consuming, it is not practically possible to annotate all of them. Therefore, we gathered a pixel-level annotation dataset such that it is minimal yet sufficient for the development of our model. Our localization dataset includes 4,814 mammograms.

\subsection{Statistical analysis}
The algorithm's classification performance was assessed using AUC, sensitivity, specificity, PPV, and NPV. The ROC curve was plotted using true-positive rate (TPR) and false-positive rate (FPR) values to determine the area under the curve (AUC). AUC reflects the overall diagnostic performance of our model. Clopper-Pearson method was used for estimating 95\% CIs for AUC, sensitivity, specificity, PPV, and NPV statistics.

Similarly, we analyze the localization performance of the algorithm by comparing the heatmap output from the model with the localization ground truths. Pixels from the predicted heatmap are assessed for true-positive or false-positive.

We measure lesion localization fraction (LLF) and non-lesion localization fraction (NLF) to assess the model's localization performance. LLF is a per-lesion metric reflecting lesion localization sensitivity that measures the ratio of true-positive localizations to total ground truth lesions in the dataset \cite{chakraborty2008operating}. NLF is a per-image metric measuring the rate of incorrect lesion highlights, representing the false-positive lesion rate in the dataset.

\subsection{Model Details}
\begin{figure*}[htbp]
    \centering
    \includegraphics[width=\linewidth]{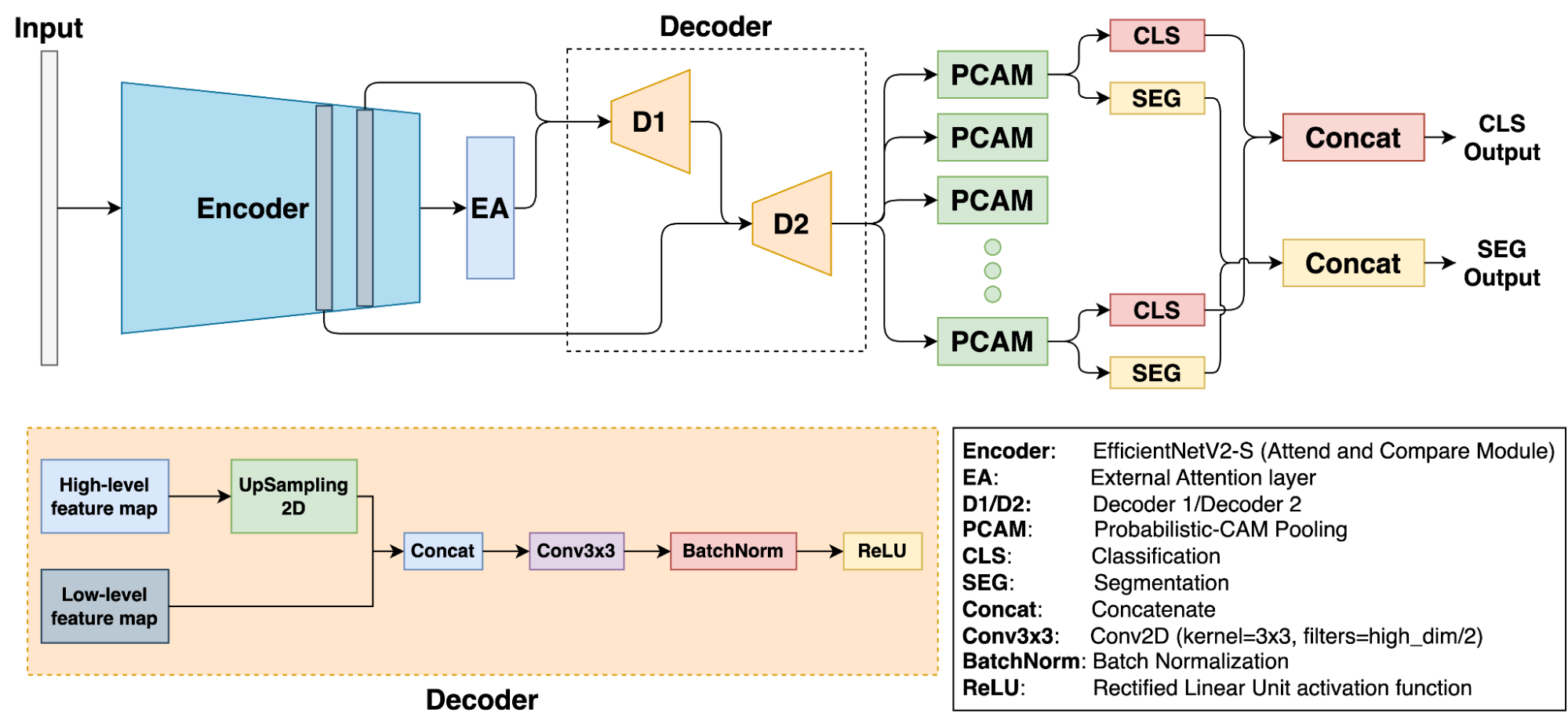}
    \caption{Architecture of the MMG model.}
    \label{fig:mmg-architecture}
\end{figure*}

Figure \ref{fig:mmg-architecture} illustrates the architecture of our deep learning model. Our deep learning model consists of three parts: an encoder, a decoder, and aggregation layers. The encoder network is based on the EfficientNetV2(S) \cite{tan2021efficientnetv2} architecture, the decoder module uses upsampling layers to increase the size of input feature maps from the encoder, and the aggregation layer applies Probabilistic-Class Activation Map or PCAM \cite{ye2020weakly} pooling to combine final feature maps to classification and heatmap outputs.

In the EfficientNetV2(S) model we replace the default Squeeze and Excitation (SE) attention mechanism with the Attend-and-Compare Module (ACM) \cite{kim2020learning}. The ACM module performs better in terms of object detection or image segmentation problems since it compares different parts of input. The SE mechanism compares the different parts of input with itself which may not be desired for object detection or image segmentation tasks. Moreover, ACM compares an object or region of interest with a corresponding context, which is similar to the way radiologists use to read X-rays. Similar to the radiologists looking at the symmetric or semantically related zones in the X-rays to compare and identify abnormalities, the ACM module also compares regions of interest with the relative context and uses the result to enhance the original image feature maps.

The External Attention (EA) layer, which learns the correlation between all data samples, is added on top of the EfficientNetV2(S) model. The output of the EA layer is forwarded to a simple decoder which increases the size of the feature map to 192x128. The scaled up feature map aids in better classification and localization of very small and scattered lesions such as "calcification". Similar to the U-Net architecture, the decoder combines high-level and low-level features by using skip connections. Finally, the output of the decoder is passed through an aggregation layer to produce two outputs, one for classification and other for segmentation (heatmap).

PCAM, a weighted average feature pooling method derived from the process of combining feature maps and probability maps (also known as Attention maps), is used to aggregate features from the underlying encoder and decoder. Spatial features, which are used to generate heatmaps identifying locations of lesions (SEG), are computed from the convolution layer without flattening the feature map from the PCAM layer, whereas for classification output (CLS), the features from PCAM layer are flattened and aggregated to generate the classification score. This allows the model to leverage both classification and segmentation loss at the same time.

A multi-task training approach is used to train the model. We train the model for classification of images as well as localization of abnormal lesions simultaneously. Our dataset contains image-level labels for classification and pixel-level annotation for localization. Pixel-level dataset aids the model in better locating and highlighting relevant areas while excluding unrelated regions.

Our training method enables models to learn from a substantial pool of partially labeled data, resulting in significant reduction in the amount of labeled data required to achieve satisfactory performance. The model is trained to classify nine different lesions.
\begin{enumerate}
    \item Suspicious Calcification
    \item Suspicious Mass
    \item Suspicious Axillary Adenopathy
    \item Suspicious Architectural Distortion
    \item Benign Calcification
    \item Benign Mass
    \item Benign Axillary Adenopathy
    \item Benign Architectural Distortion
    \item Normal
\end{enumerate}

The lesions detected by the model are subsequently classified into three main categories: Calcification, Mass, and Other. The detection of these lesions is based on the combined output from multiple nodes within the model as shown in Table \ref{tab:final-categories}. In cases where the model identifies multiple suspicious lesions within each final category, the lesion with the highest probability score is selected for further consideration.

Benign lesions are also taken into account during the calculation process. However, to ensure that radiologists are aware of these lesions without being excessively focused on them, the maximum probability for benign lesions is limited to 15\%. This approach allows medical professionals to note the presence of benign lesions while prioritizing their attention on potentially more significant findings.

\begin{table}[htbp]
    \centering
    \caption{Final categories of output classes after combining different output nodes from the model.}
    \label{tab:final-categories}
    \begin{tabular}{p{0.2\linewidth} p{0.7\linewidth}}
        \toprule
        \textbf{Final Category} & \textbf{Corresponding output nodes} \\
        \midrule
        Calcification & Calcification (Suspicious only) \\
        Mass & Mass\\
        Other & Axillary Adenopathy, Architectural Distortion \\        
        \bottomrule
    \end{tabular}
\end{table}

The model takes input images of 1024 x 1536 (width x height) pixels. Images are preprocessed before training the model. Following steps are used for preprocessing.
\begin{enumerate}
    \item Convert 14-bit DICOM format mammogram images to 8-bit png format.
    \item Filter out the background of the image by discarding pixels with intensity value less than 10.
    \item Filter out and remove the LCC, LMLO, RCC and RMLO view tags in the image using the connected components algorithm \cite{opencv-connected}. This segmentation process will isolate only the breast region of interest in the image.
    \item Occasionally, the breast region from step 3 might miss some parts within the breast area resulting in incomplete breast image. To overcome this problem, we first create a breast mask from the largest component in step 3, then the missing regions are filled using image dilation and erosion techniques. This creates a single blob of the breast region which is later used to only keep the breast area while removing everything beyond it. This step ensures that unwanted artifacts such as mammographic view tags like LCC, RMLO, etc. present in the image are removed and only the breast part is retained.
    \item Finally, the images are resized to 1024 x 1536 (width x height) pixels
\end{enumerate}

\begin{figure}[htbp]
    \centering
    \begin{subfigure}{0.45\linewidth}
        \includegraphics[width=\linewidth]{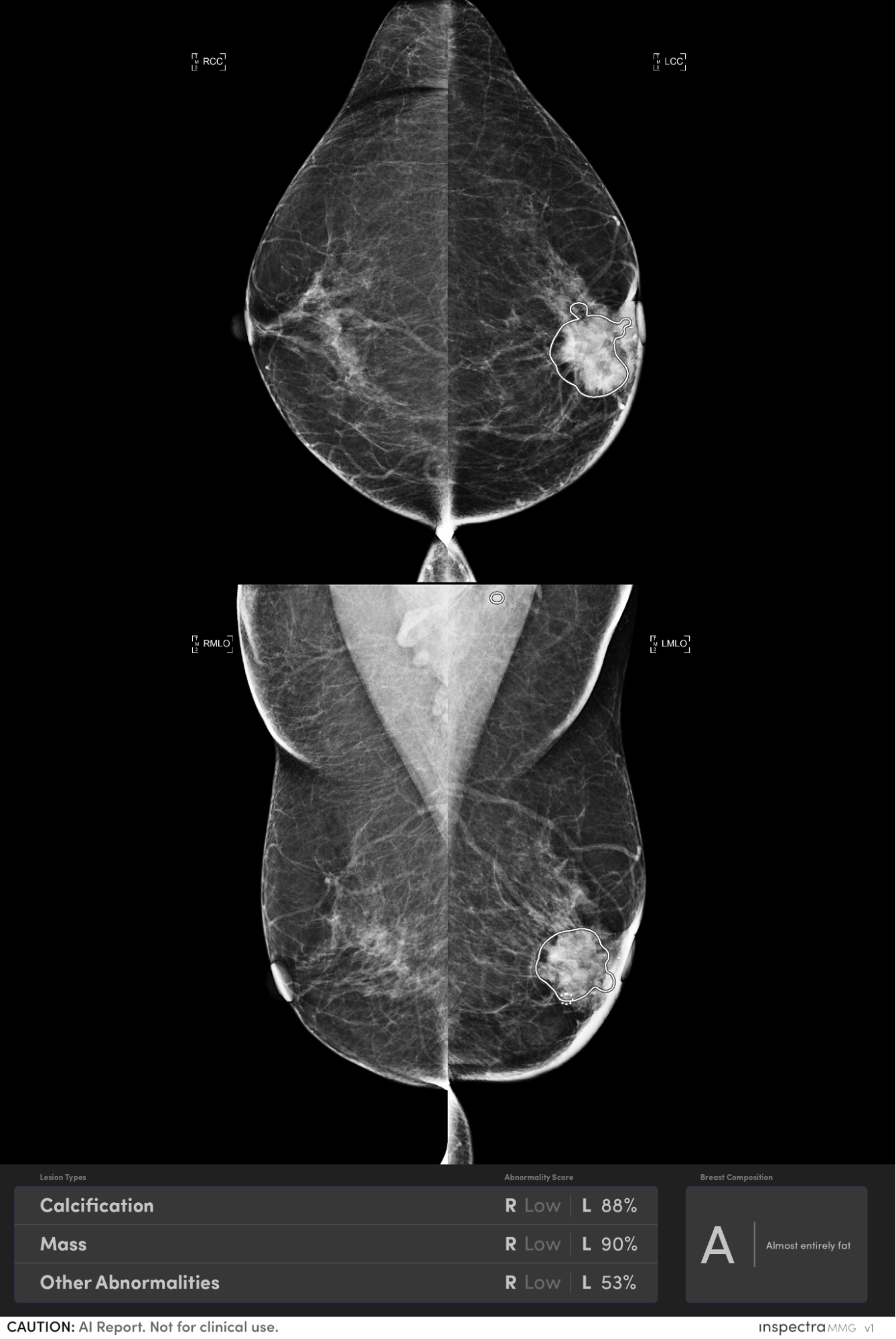}
        \caption{Greyscale heatmap.}
    \end{subfigure}
    \begin{subfigure}{0.45\linewidth}
        \includegraphics[width=\linewidth]{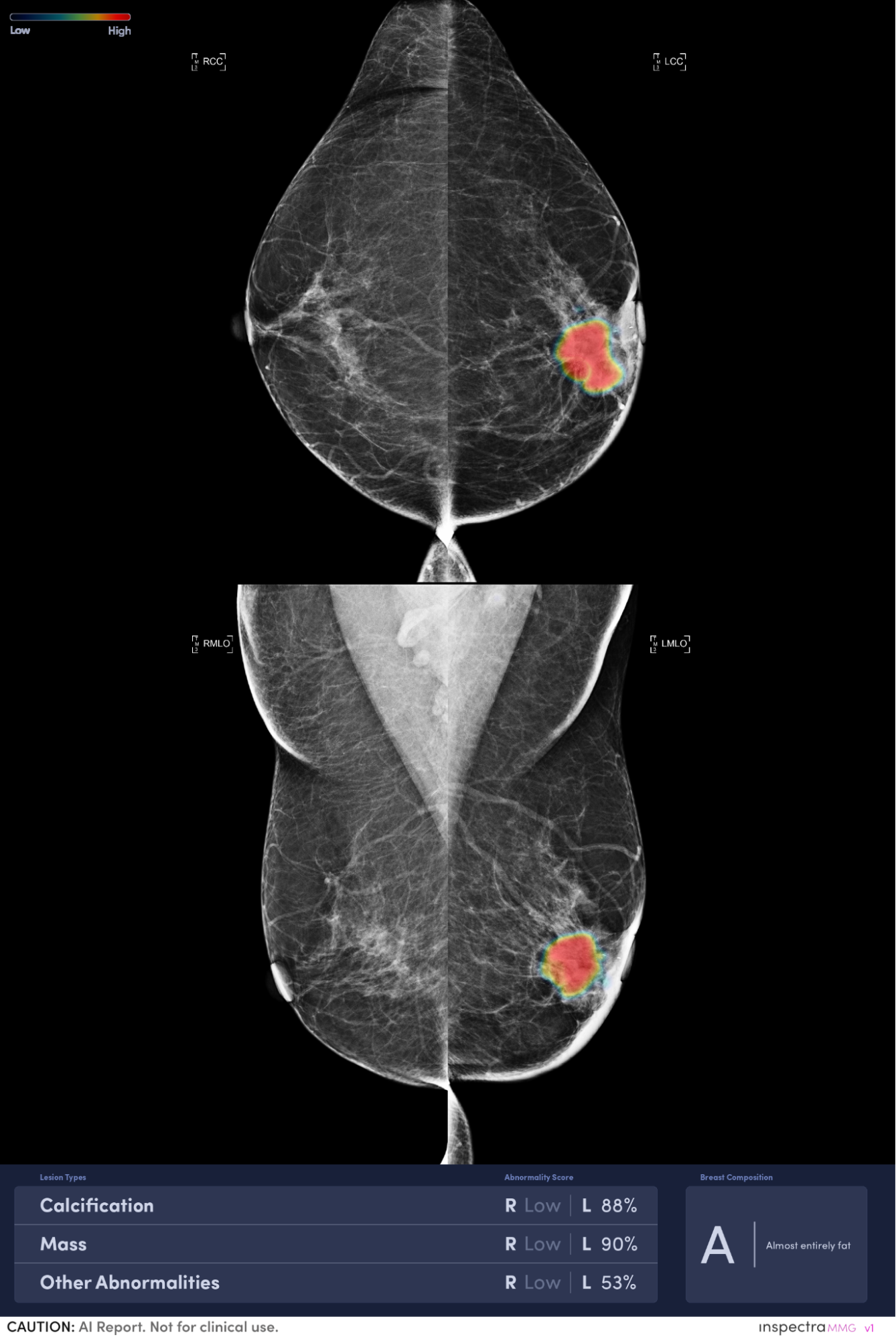}
        \caption{Color heatmap.}
    \end{subfigure}
    \caption{Final output from the MMG model.}
    \label{fig:model-output}
\end{figure}

A confidence score and heatmap for each class is estimated by the model. Suspicion score per class (Calcification, Mass, Other) by breast (left/right) is presented to the end users. Heatmaps suggesting the location of suspicious lesions are also overlaid on the final image. Heatmaps are also generated for benign lesions and are visually represented using distinct styles to subtly alert users to their presence. In color heatmaps, benign lesions are depicted using a blue color scheme, while in greyscale heatmaps, they are represented by dotted lines. These visual cues aim to draw the user's attention to the benign lesions without overemphasizing their importance. Figure \ref{fig:model-output} shows examples of the final output images with confidence scores and overlaid heatmaps.

\section{Clinical Evaluation}
We perform clinical evaluation on two aspects: classification and localization as mentioned in Section: Statistical Analysis. For each aspect, we assess the performance on three different datasets - (1) in-domain mammograms which are not seen by the model before, (2) mammograms that are pathologically confirmed for cancer, and (3) out-of-domain generalizability set from other hospitals. This study was approved by the ethics review and institutional review board from participating institutions, and the requirement for informed consent was waived.

The in-domain test set is a part of the Siriraj dataset kept aside for evaluation of the model. Pathologically or biopsy-confirmed dataset includes mammogram-guided biopsy cases and two-year follow-up confirmed benign and normal samples. For assessing out-of-domain generalizability, we used a diverse dataset collected from various hospitals. Detailed information about these datasets is provided in Table \ref{tab:validation-sets} and Table \ref{tab:localization-labels}.

\begin{table}[htbp]
    \centering
    \caption{Distribution of different validation sets used for model evaluation.}
    \label{tab:validation-sets}
    \begin{tabular}{\datasettable}
        \toprule
        \textbf{Category} & \textbf{In-domain test set} & \textbf{Biopsy-confirmed set} & \textbf{Generalization set} \\
        \midrule
        \textbf{Overall dataset size} & & & \\
        Total Cases & 9,421 & 883 & 761 \\
        \midrule
        \textbf{Breast composition} & & & \\
        Almost entirely fat & 141 & 20 & 19 \\
        Scattered areas of fibroglandular densities & 2,012 & 227 & 113 \\
        Heterogeneously dense & 6,722 & 601 & 432 \\
        Extremely dense & 546 & 35 & 48 \\
        \midrule
        \textbf{BIRADS Category} & & & \\
        1 & 1,448 & 298 & 195 \\
        2 & 6,500 & 201 & 197 \\
        3 & 315 & - & 2 \\
        4 & 914 & 167 & 154 \\
        5 & 177 & 207 & 181 \\
        6 & 67 & 10 & 28 \\
        \midrule
        \textbf{Abnormality type} & & & \\
        Malignant or Suspicious & 1,473 & 385 & 371 \\
        Benign & 6,500 & 200 & 195 \\
        Normal & 1,448 & 298 & 195 \\
        \midrule
        \bottomrule
    \end{tabular}
\end{table}

\begin{table}[htbp]
    \centering
    \caption{Number of lesions with localization ground truth labels in the validation dataset.}
    \label{tab:localization-labels}
    \begin{tabular}{\datasettable}
        \toprule
        \textbf{Lesion type} & \textbf{In-domain test set} & \textbf{Biopsy-confirmed set} & \textbf{Generalization set} \\
        \midrule
        Calcification & 497 & 392 & 325 \\
        Mass & 557 & 569 & 630 \\
        Other & 863 & 188 & 193 \\
        \bottomrule
    \end{tabular}
\end{table}

\subsection{In-domain evaluation methodology and results}
As shown in Figure \ref{fig:mmg-development}, our in-domain validation set contains 9,421 mammograms. Table \ref{tab:detection-performance} summarizes the classification performance of our model on in-domain dataset. With an AUROC of above 0.89 and sensitivity of over 87\% across all three conditions, our model demonstrated impressive performance on the test set. The AUROC reflects superior classification performance across different abnormal conditions in mammography. All values are computed based on the optimal operating point of the model.

The model demonstrates superior cancer detection ability reflected by the AUC of above 0.89 and sensitivity of 93\% and NPV of 98\%, making it suitable for mammogram screening process. The lower specificity of 58\% on the in-domain test set results from its higher proportion of benign (BI-RADS 2) samples. The model struggles more with benign lesions than normal tissue (BI-RADS 1), producing more false positives as benign findings share some visual features with malignant cases while still being non-malignant.

\begin{figure}[htbp]
    \centering
    \includegraphics[width=0.7\linewidth]{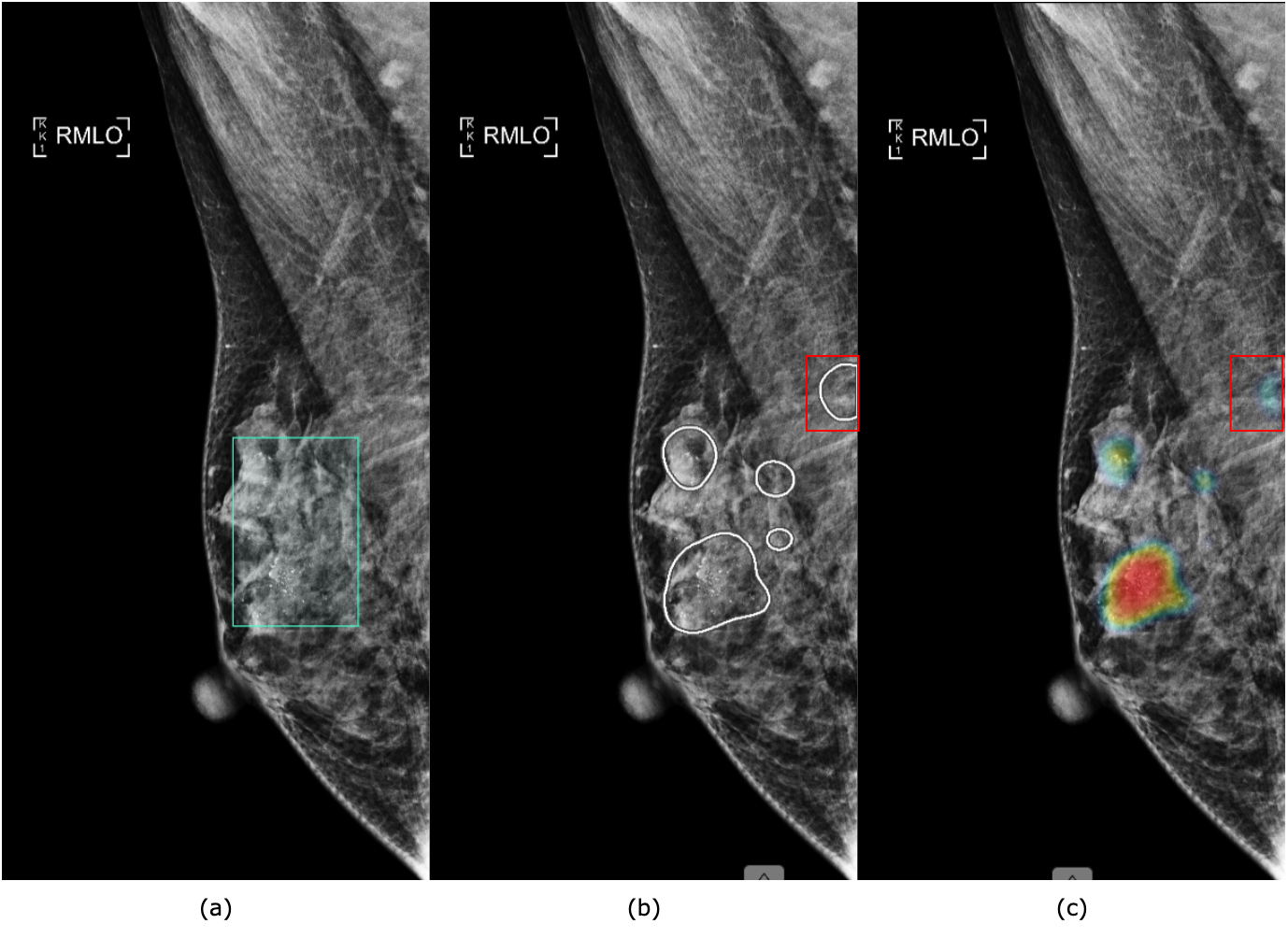}
    \caption{An RMLO view of a patient positive for Calcification. (a) Ground truth location of calcification in the breast which was verified by expert radiologists. (b) Predicted heatmap by the model in a monochrome output setting. The polygons in white solid boundaries are the heatmap output from the model. The polygon within the red box is a false-positive prediction by the model, all other polygons are true-positive heatmaps. (c) Predicted heatmap by the model in a color output setting. The blob within the red box is a false-positive heatmap.}
    \label{fig:rmlo-view}
\end{figure}

Our model achieved an impressive performance in lesion localization as seen in Table \ref{tab:localization-performance}. A true-positive rate (LLF) of around 74 for mass and calcification indicates that the model accurately identifies the location of most lesions in the test set. LLF of 60 for other lesions indicates that the model does fairly well in locating axillary adenopathy and/or architectural distortion as well. Additionally, an NLF of 0.81 or less suggests that, on average, there are fewer than 0.81 false heatmaps per case. Along with high average IoGT and IoHM values of 0.31 and 0.37, these metrics demonstrate that our model effectively locates suspicious lesions while keeping the number of false highlights relatively low. Figure \ref{fig:rmlo-view} shows both the true-positive and false-positive heatmaps from our model.

\begin{table}[htbp]
    \centering
    \caption{Detection performance on the in-domain test set. \cancerexplain}
    \label{tab:detection-performance}
    \begin{tabular}{\perftable}
        \toprule
        \textbf{Condition} & \textbf{AUROC} & \textbf{PPV} & \textbf{NPV} & \textbf{Sens} & \textbf{Spec} \\
        \midrule
        Cancer & 0.894 & 0.289 & 0.978 & 0.929 & 0.581 \\
        & (0.883 - 0.905) & (0.276 - 0.302) & (0.974 - 0.982) & (0.915 - 0.942) & (0.570 - 0.592) \\
        \midrule
        Calcification & 0.936 & 0.325 & 0.988 & 0.885 & 0.841 \\
        & (0.923 - 0.948) & (0.304 - 0.345) & (0.986 - 0.991) & (0.860 - 0.907) & (0.834 - 0.849) \\
        Mass & 0.950 & 0.178 & 0.992 & 0.896 & 0.760 \\
        & (0.936 - 0.963) & (0.164 - 0.194) & (0.990 - 0.994) & (0.866 - 0.921) & (0.751 - 0.769) \\
        Other & 0.890 & 0.204 & 0.990 & 0.869 & 0.790 \\
        & (0.871 - 0.908) & (0.188 - 0.221) & (0.987 - 0.992) & (0.838 - 0.896) & (0.781 - 0.798) \\
        \bottomrule
    \end{tabular}
\end{table}

\begin{table}[htbp]
    \centering
    \caption{Localization performance on in-domain test set. \cancerexplain}
    \label{tab:localization-performance}
        \begin{tabular}{\localizationtable}
            \toprule
            \textbf{Condition} & \textbf{\#Lesions} & \textbf{LLF} & \textbf{NLF} \\
            \midrule
            Cancer & 1,879 & 0.712 & 1.66 \\
            & & (0.691 - 0.732) & (1.634 - 1.686) \\            
            \midrule
            Calcification & 497 & 0.743 & 0.81 \\
            & & (0.702 - 0.780) & (0.791 - 0.828) \\
            Mass & 557 & 0.734 & 0.40 \\
            & & (0.696 - 0.771) & (0.389 - 0.415) \\
            Other & 863 & 0.604 & 0.46 \\
            & & (0.570 - 0.637) & (0.444 - 0.471) \\
            \bottomrule
        \end{tabular}
\end{table}

\subsection{Evaluation on biopsy-confirmed cancer dataset}
We acquired another set of mammography dataset from Thanyarak Breast Center at Siriraj Hospital in Bangkok, Thailand, which included cases of cancer that were pathologically confirmed. In 2022, biopsies were carried out on these patients, using both ultrasonography and mammography guidance. Since our model was developed using mammography images, we discarded biopsy cases that were guided by ultrasound for a fair evaluation of our model. In addition to cancer confirmed cases, we included benign and normal samples confirmed by 2 years of follow up in this dataset. The final dataset consists of 883 cases as mentioned in Table \ref{tab:validation-sets}. Cancer ground truths were established with the help of expert radiologists referring to both radiology reports and pathology reports. Individual case-level classification labels and lesion-level localization boundaries were verified by expert radiologists.

Our algorithm demonstrated excellent performance in the biopsy-confirmed set. Table \ref{tab:biopsy-detection} and Table \ref{tab:biopsy-localization} summarizes the performance on the biopsy-confirmed set. Our model is highly sensitive to cancer suspicious lesions as shown by the sensitivity values of at-least 88\% in Table \ref{tab:biopsy-detection}. The model also demonstrates an impressive distinction between multiple lesion types as shown by the AUC of around 0.90 in every lesion class. Our model achieved excellent results, AUC above 0.96, in final cancer classification.

Table \ref{tab:biopsy-localization} presents our model's localization performance on biopsy-confirmed cancer lesions. The model achieved LLF values of above 80 for calcification and mass classes, indicating good accuracy in highlighting these types of lesions. For the "other" class, an LLF of approximately 70 suggests that the model locates these lesions with good precision. Additionally, the low NLF score of less than one indicates that the model won't cause unnecessary distractions for reading radiologists.

\begin{table}[htbp]
    \centering
    \caption{Detection performance on the biopsy-confirmed dataset. \cancerexplain}
    \label{tab:biopsy-detection}
    \begin{tabular}{\perftable}
        \toprule
        \textbf{Condition} & \textbf{AUROC} & \textbf{PPV} & \textbf{NPV} & \textbf{Sens} & \textbf{Spec} \\
        \midrule
        Cancer & 0.963 & 0.759 & 0.967 & 0.966 & 0.763 \\
        & (0.950 - 0.977) & (0.719 - 0.796) & (0.944 - 0.982) & (0.943 - 0.982) & (0.723 - 0.800) \\
        \midrule
        Calcification & 0.968 & 0.691 & 0.976 & 0.921 & 0.888 \\
        & (0.951 - 0.986) & (0.629 - 0.747) & (0.961 - 0.987) & (0.872 - 0.955) & (0.862 - 0.910) \\
        Mass & 0.963 & 0.659 & 0.956 & 0.920 & 0.785 \\
        & (0.947 - 0.979) & (0.609 - 0.706) & (0.934 - 0.972) & (0.881 - 0.949) & (0.750 - 0.817) \\
        Other & 0.894 & 0.333 & 0.978 & 0.882 & 0.749 \\
        & (0.854 - 0.934) & (0.279 - 0.391) & (0.963 - 0.988) & (0.806 - 0.936) & (0.717 - 0.779) \\
        \bottomrule
    \end{tabular}    
\end{table}

\begin{table}[htbp]
    \centering
    \caption{Localization performance on the biopsy-confirmed dataset. \cancerexplain}
    \label{tab:biopsy-localization}
    \begin{tabular}{\localizationtable}
        \toprule
        \textbf{Condition} & \textbf{\#Lesions} & \textbf{LLF} & \textbf{NLF} \\
        \midrule
        Cancer & 928 & 0.861 & 1.28 \\
        & & (0.837 - 0.883) & (1.206 - 1.357) \\
        \midrule
        Calcification & 392 & 0.847 & 0.37 \\
        & & (0.807 - 0.881) & (0.331 - 0.413) \\
        Mass & 569 & 0.810 & 0.71 \\
        & & (0.776 - 0.842) & (0.656 - 0.768) \\
        Other & 189 & 0.697 & 0.42 \\
        & & (0.626 - 0.762) & (0.378 - 0.465) \\
        \bottomrule
    \end{tabular}
\end{table}

\subsection{Evaluation on out-of-domain dataset}

To evaluate the generalizability of our model, we collected a dataset from a network of hospitals different from our development dataset. This dataset comprises 761 breast mammography images, as detailed in Table \ref{tab:validation-sets}. Despite its relatively small size, this dataset is instrumental in assessing the model's performance on out-of-domain data.

Table \ref{tab:gen-detection} and Table \ref{tab:gen-localization} shows the performance on the generalization set. With an AUC of 0.937, our model demonstrates excellent performance on this dataset as well. The overall sensitivity and specificity of 93.0\% and 68.7\% for cancer detection shows a slight decrease but remains clinically acceptable, indicating robust generalizability across different clinical environments.

The model's localization performance on out-of-domain data (Table \ref{tab:gen-localization}) shows varying results across lesion types. While calcification detection maintained strong performance (LLF 0.788), mass localization showed some degradation (LLF 0.657), suggesting that mass appearance may vary more significantly across different hospital settings due to differences in imaging protocols or patient populations. The relatively small decrease in overall performance metrics compared to the biopsy-confirmed dataset demonstrates the model's ability to maintain clinical utility when deployed in new healthcare settings, though initial monitoring of performance during deployment would be advisable.

\begin{table}[htbp]
    \centering
    \caption{Detection performance on the generalizability dataset. \cancerexplain}
    \label{tab:gen-detection}
    \begin{tabular}{\perftable}
        \toprule
        \textbf{Condition} & \textbf{AUROC} & \textbf{PPV} & \textbf{NPV} & \textbf{Sens} & \textbf{Spec} \\
        \midrule
        Cancer & 0.937 & 0.739 & 0.912 & 0.930 & 0.687 \\
        & (0.918 - 0.955) & (0.696 - 0.778) & (0.873 - 0.941) & (0.899 - 0.954) & (0.639 - 0.733) \\
        \midrule
        Calcification & 0.948 & 0.663 & 0.960 & 0.863 & 0.884 \\
        & (0.924 - 0.972) & (0.595 - 0.727) & (0.940 - 0.975) & (0.799 - 0.912) & (0.855 - 0.908) \\
        Mass & 0.927 & 0.759 & 0.850 & 0.810 & 0.807 \\
        & (0.907 - 0.948) & (0.710 - 0.803) & (0.812 - 0.883) & (0.763 - 0.851) & (0.767 - 0.843) \\
        Other & 0.786 & 0.305 & 0.950 & 0.802 & 0.671 \\
        & (0.735 - 0.837) & (0.254 - 0.360) & (0.925 - 0.968) & (0.717 - 0.870) & (0.634 - 0.707) \\
        \bottomrule
    \end{tabular}
\end{table}

\begin{table}[htbp]
    \centering
    \caption{Localization performance on the generalizability dataset. \cancerexplain}
    \label{tab:gen-localization}
    \begin{tabular}{\localizationtable}
        \toprule
        \textbf{Condition} & \textbf{\#Lesions} & \textbf{LLF} & \textbf{NLF} \\
        \midrule
        Cancer & 784 & 0.796 & 1.38 \\
        & & (0.766 - 0.824) & (1.295 - 1.463) \\
        \midrule
        Calcification & 325 & 0.788 & 0.53 \\
        & & (0.739 - 0.831) & (0.482 - 0.587) \\
        Mass & 630 & 0.657 & 0.28 \\
        & & (0.619 - 0.694) & (0.247 - 0.324) \\
        Other & 193 & 0.461 & 0.56 \\
        & & (0.389 - 0.534) & (0.507 - 0.614) \\
        \bottomrule
    \end{tabular}    
\end{table}

\section{System Usability}
To evaluate the reliability of Inspectra MMG model and determine the extent to which radiologists accept its results in clinical practice, we conducted a series of evaluations. Specifically, We conducted a study to evaluate the concordance between radiology reports and the results of Inspectra MMG model in detecting and localizing abnormalities within a pathologically confirmed malignant dataset \cite{kim2022concordance}. Additionally, we presented the AI results to breast specialists and measured the level of acceptance of the results when using them in a clinical setting \cite{gaube2021do}. Finally, we conducted a survey to assess the usability of the system by using the system usability scale \cite{brooke2013sus}.

\subsection{Concordance Test}
Our concordance study was conducted on the biopsy confirmed set as mentioned in Table \ref{tab:validation-sets}. To evaluate whether the findings from the model were reflected in the radiology reports, we divided the concordance study into two parts: classification concordance test and localization concordance test.

For classification concordance, cases where both the radiology reports and the model results were negative were classified as "accept." Cases where the radiology reports were positive but the model was negative, or vice versa, were classified as "reject." For cases where both the radiology reports and the model results were positive, classifications were made based on lesion concordance: "accept" if the lesions described in both were in agreement, "edit" if the lesions partially agreed, and "add" if the radiology reports identified additional lesions not detected by the model. If the lesion described in the radiology report differed from that detected by the model, the case was classified as "reject." The classifications "accept," "edit," and "add" represented concordance, while "reject" indicated discordance. The concordance rate was then evaluated based on the lesion categories.

For localization concordance, heatmaps from the MMG model were compared with the localization ground truths provided by the radiologists and then divided into one of the four categories based on the agreements between them, as shown in Table \ref{tab:biopsy-detection}. The  MMG model heatmaps were presented to the radiologists as a complete representation of the localization of all lesions, without considering the specific lesion types. While previous work \cite{kim2022concordance} focused on classification concordance, we have extended this approach to include localization results as well. The detailed criteria for each concordance type are shown in Table \ref{tab:categorization-rules} and Table \ref{tab:heatmap-rules}.

\begin{table}[htbp]
    \centering
    \caption{Criteria for Inspectra MMG Model Classification Concordance.}
    \label{tab:categorization-rules}
    \begin{tabular}{p{0.15\linewidth} | p{0.55\linewidth} | p{0.15\linewidth}}
        \toprule
        \textbf{Category} & \textbf{Criteria} & \textbf{Result Category} \\
        \midrule
        Agree & Radiologist report completely matches with AI. & Concordant \\
        Edit & Radiologist report is in partial agreement with AI. & Concordant \\
        Add & Radiologist report added some conditions to the AI results. & Concordant \\
        Reject & Radiologist reports and AI have different conditions. & Discordant \\
        \bottomrule
    \end{tabular}
\end{table}

\begin{table}[htbp]
    \centering
    \caption{Criteria for Inspectra MMG Model Localization Concordance.}
    \label{tab:heatmap-rules}
    \begin{tabular}{p{0.15\linewidth} | p{0.55\linewidth} | p{0.15\linewidth}}
        \toprule
        \textbf{Category} & \textbf{Criteria} & \textbf{Result Category} \\
        \midrule
        Agree & Ground-truth box overlap with AI heatmap more than 50\% on all lesions & Concordant \\
        Edit & AI heatmap covers all lesion areas but also highlight some non-lesion areas & Concordant \\
        Add & AI heatmap misses some lesions & Concordant \\
        Reject & No overlap between AI heatmap and lesion areas & Discordant \\
        \bottomrule
    \end{tabular}
\end{table}

Our study analyzed concordance rates across two datasets. Table \ref{tab:biopsy-concordance} shows results from the biopsy-confirmed dataset, where of 883 samples, the classification concordance rate was 83.47\% (agree: 59.23\%, edit: 21.86\%, add: 2.38\%) and discordance rate was 16.53\%. Similarly, the localization concordance rate was 84.03\% (agree: 52.66\%, edit: 23.56\%, add: 7.81\%) and discordance rate was 15.97\%.

For the generalizability dataset shown in Table \ref{tab:gen-concordance}, of 761 samples, the classification concordance rate was 78.1\% (agree: 51.2\%, edit: 19.6\%, add: 7.2\%) and discordance rate was 21.9\%. The localization concordance rate was 79.6\% (agree: 49.4\%, edit: 23.4\%, add: 6.8\%) and discordance rate was 20.4\%.

These results demonstrate strong performance of Inspectra MMG model across both datasets. The comparable concordance rates between the biopsy-confirmed dataset and the generalizability dataset (classification: 83.47\% vs. 78.1\%; localization: 84.03\% vs. 79.6\%) suggest that our model maintains robust performance when applied to new data. The relatively small difference in concordance rates indicates that the model generalizes well beyond the initial biopsy-confirmed cases, which is particularly promising for real-world clinical applications.

\begin{table}[htbp]
    \centering
    \caption{Concordance rate between radiologists' reports and Inspectra MMG model on the biopsy-confirmed dataset}
    \label{tab:biopsy-concordance}
    \begin{tabular}{p{0.25\linewidth} | p{0.10\linewidth} | p{0.15\linewidth} | p{0.10\linewidth} | p{0.15\linewidth}}
        \toprule
        \multirow{2}{*}{\textbf{Category}} & \multicolumn{2}{c}{\textbf{Classification}} & \multicolumn{2}{c}{\textbf{Localization}} \\
        \cmidrule(lr){2-3} \cmidrule(lr){4-5}
        & \textbf{\# cases} & \textbf{\%} & \textbf{\# cases} & \textbf{\%} \\
        \midrule
        Agree & 523 & 59.23\% & 465 & 52.66\% \\
        Edit & 193 & 21.86\% & 208 & 23.56\% \\
        Add & 21 & 2.38\% & 69 & 7.81\% \\
        Reject & 146 & 16.53\% & 141 & 15.97\% \\
        \midrule
        Concordant & 737 & 83.5\% (0.808 - 0.859) & 742 & 84.0\% (0.814 - 0.864) \\
        \bottomrule
    \end{tabular}
\end{table}

\begin{table}[htbp]
    \centering
    \caption{Concordance rate between radiologists' reports and Inspectra MMG model on the generalizability dataset}
    \label{tab:gen-concordance}
    \begin{tabular}{p{0.25\linewidth} | p{0.10\linewidth} | p{0.15\linewidth} | p{0.10\linewidth} | p{0.15\linewidth}}
        \toprule
        \multirow{2}{*}{\textbf{Category}} & \multicolumn{2}{c}{\textbf{Classification}} & \multicolumn{2}{c}{\textbf{Localization}} \\
        \cmidrule(lr){2-3} \cmidrule(lr){4-5}
        & \textbf{\# cases} & \textbf{\%} & \textbf{\# cases} & \textbf{\%} \\
        \midrule
        Agree & 390 & 51.2\% & 376 & 49.4\% \\
        Edit & 149 & 19.6\% & 178 & 23.4\% \\
        Add & 55 & 7.2\% & 52 & 6.8\% \\
        Reject & 167 & 21.9\% & 155 & 20.4\% \\
        \midrule
        Concordant & 594 & 78.1\% (0.749 - 0.809) & 606 & 79.6\% (0.766 - 0.824) \\
        \bottomrule
    \end{tabular}
\end{table}

\subsection{Usefulness of the MMG Model}
To further evaluate the usefulness of Inspectra MMG model results in a clinical setting, we computed the average acceptance score across all radiologists participating in the study. The acceptance score of a radiologist is the ratio of the number of accepted cases by the radiologist to total cases in the study. This score is computed for each radiologist and the mean score across all radiologists is reported as the final acceptance score.

We presented the prediction scores and heatmaps to the radiologists and asked them to grade the results in the score range of 1-4, 1: not useful, 2: neutral, 3: useful and 4: extremely useful. A score of 2 or above is considered acceptable.

Two expert radiologists reviewed each case's results from AI and compared them with the ground truth annotations. Based on the correctness of classification and localization results, the radiologist rated each case.

For the biopsy-confirmed dataset, considering that the radiologists were required to review all 883 cases, the process would have demanded substantial time and effort. To mitigate their workload, we excluded the 423 cases which were categorized as 'Agree' in both classification and localization concordance tests. These cases are considered as 'Accept' in the final acceptance score calculation since they represent cases that are correct in both classification and localization.

The radiologists evaluated the remaining 460 cases categorized as 'Add', 'Edit', and 'Reject' from the classification and localization concordance tests. Table \ref{tab:radiologist-scores-biopsy} outlines the average number of cases accepted by the radiologists. On average, the radiologists accepted 430.5 cases (in addition to the 423 automatically accepted cases), resulting in a final average acceptance rate of 96.7\% out of 883 cases.

For the generalization dataset shown in Table \ref{tab:radiologist-scores-gen}, we followed a similar approach. Of the 761 total cases, 307 cases categorized as 'Agree' in both classification and localization were automatically considered as 'Accept'. Radiologists evaluated the remaining cases, accepting a total of 679.5 cases (including the automatic accepts), resulting in an acceptance rate of 89.3\%.

These high acceptance rates across both datasets indicate that the radiologists were satisfied with the performance of Inspectra MMG model in clinical practice. The consistent performance between the biopsy-confirmed dataset (96.7\% acceptance) and the generalization dataset (89.3\% acceptance) demonstrates that our model maintains clinical utility when applied to new data. This relatively small difference in acceptance rates suggests robust generalizability, which is crucial for the model's practical implementation in diverse clinical settings.

\begin{table}[htbp]
    \centering
    \caption{Radiologist's scores for the reviewed cases on the biopsy-confirmed dataset. Cases where AI and radiologists agree in both classification and localization results are automatically accepted. AA stands for automatically accepted cases that receive agree concordance in both classification and localization test.}
    \label{tab:radiologist-scores-biopsy}
    \begin{tabular}{lcc}
        \toprule
        \textbf{Score} & \textbf{Acceptance} & \textbf{Average Count} \\
        \midrule
        1 & Not Accept & 29.5 \\
        2 & Accept & 84 \\
        3 & Accept & 290.5 \\
        4 & Accept & 56 \\
        AA & Accept & 423 \\
        \midrule
        \multicolumn{2}{l}{Acceptance Rate} & 96.7\% (0.953 - 0.977) \\
        \bottomrule
    \end{tabular}
\end{table}

\begin{table}[htbp]
    \centering
    \caption{Radiologist's scores for the reviewed cases on the generalization dataset. Cases where AI and radiologists agree in both classification and localization results are automatically accepted. AA stands for automatically accepted cases that receive agree concordance in both classification and localization test.}
    \label{tab:radiologist-scores-gen}
    \begin{tabular}{lcc}
        \toprule
        \textbf{Score} & \textbf{Acceptance} & \textbf{Average Count} \\
        \midrule
        1 & Not Accept & 81.5 \\
        2 & Accept & 223.5 \\
        3 & Accept & 123 \\
        4 & Accept & 26 \\
        AA & Accept & 307 \\
        \midrule
        \multicolumn{2}{l}{Acceptance Rate} & 89.3\% (0.869 - 0.914) \\
        \bottomrule
    \end{tabular}
\end{table}

\subsection{System Usability Test}

Furthermore, a questionnaire to assess the System Usability Scale (SUS) \cite{brooke2013sus} score was distributed to radiologists with at least one month of Inspectra MMG usage experience. Assessments were conducted at the source hospital where training data was collected and at validation hospitals with different patient populations and imaging protocols.

Table \ref{tab:sus-score} shows the SUS scores from both hospital types. For the source hospital, 18 certified breast specialist radiologists who had used Inspectra MMG for at least one month participated. The model achieved an average SUS score of 74.17.
For validation hospitals, 22 certified breast specialist radiologists participated in the usability assessment, achieving an average SUS score of 69.20.
The SUS scores (74.17 vs 69.20) demonstrate consistent usability with only a 5-point difference, indicating the model's interface remains user-friendly across different clinical environments. These reasonable SUS scores, combined with high concordance rates and strong radiologist acceptance, suggest Inspectra MMG has potential as a reliable AI assistant for mammogram interpretation in diverse clinical settings.

\begin{table}[htbp]
    \centering
    \caption{System usability score for the MMG model from radiologists from source hospital where training data was collected and validation hospitals with different patient populations and imaging protocols.}
    \label{tab:sus-score}
    \begin{tabular}{lcc}
        \toprule
        \textbf{Hospital} & \textbf{\# Participants} & \textbf{SUS score} \\
        \midrule
        Source hospital & 18 & 74.17 (69.8 - 78.6) \\
        Validation hospitals & 22 & 69.20 (62.9 - 75.5) \\
        \bottomrule
    \end{tabular}
\end{table}

\section{Conclusion}
We presented a clinical validation study of the MMG model that encompasses a comprehensive evaluation of its performance on in-sample, biopsy-confirmed, and generalizability test sets. Furthermore, we conducted several studies, including concordance tests and usefulness tests involving expert radiologists, to assess the system's usability in real-world clinical settings.

Based on the evidence of the model's excellent performance, Inspectra MMG model demonstrates exceptional accuracy in detecting and localizing suspicious lesions across diverse datasets. These findings support the model's potential to enhance workflows in clinical practice, offering reliable assistance to radiologists in identifying malignancies. Future work may focus on further refining the model's capabilities, expanding its application to broader populations, and integrating it seamlessly into routine clinical use.

\subsection*{Ethics approval and consent to participate}
This study complied with the Declaration of Helsinki and was approved by the institutional review board from the Siriraj Institutional Review Board (Si990/2021), the Bangkok Hospital Headquarters Institutional Review Board (BHQ-IRB2024-05-21), and the Khon Kaen University Ethics Commitee for Human Research (HE674023). Informed consent was waived due to the retrospective nature of the study.

\subsection*{Consent for publication}
Not applicable

\bibliography{references}

\end{document}